\documentclass{llncs}
\usepackage{llncsdoc}
\usepackage{graphicx}
\usepackage{subfigure}
\usepackage{subfig}
\begin{document}
\thispagestyle{empty}

\newcounter{save}\setcounter{save}{\value{section}}
{\def\addtocontents#1#2{}%
\def\addcontentsline#1#2#3{}%
\def\markboth#1#2{}%
\title{A performance Analysis of the Game of Life based on parallel algorithm}

\author{Longfei Ma, Xue Chen and Zhouxiang Meng}

\institute{Computer Science and Engineering Department, Sichuan University JinJiang College,620860 Pengshan, China \\fly3601@gmail.com, smile.x.mo@gmail.com, mengzhouxiang007@gmail.com}
\maketitle
\begin{abstract}
In this article, Conway's Game of Life using OpenMP parallel processing to simulate several different parallel methods, experimental performance results and compare to find the optimal solution of the parallelization of the Game of Life. Finally pointed out the importance of the design of parallel algorithms in solving the parallel problem.
\end{abstract}

\section{Introduction}
The Game Of Life is a zero-player game, its evolution is determined by its initial state, requiring no further input. One interacts with the Game of Life by creating an initial configuration and observing how it evolves.

The universe of the Game of Life is an infinite two-dimensional orthogonal grid of square cells, each of which is in one of two possible states, alive or dead. Every cell interacts with its eight neighbors, which are the cells that are horizontally, vertically, or diagonally adjacent. At each step in time, the following transitions occur:
 \begin{enumerate}
\item Any live cell with fewer than two live neighbors dies, as if caused by under-population.
\item Any live cell with two or three live neighbors lives on to the next generation.
\item Any live cell with more than three live neighbors dies, as if by overcrowding.
\item Any dead cell with exactly three live neighbors becomes a live cell, as if by reproduction.
\end{enumerate}

OpenMP (Open Multiprocessing) is an API that supports multi-platform shared memory multiprocessing programming in C, C++, and Fortran, on most processor architectures and operating systems, including Solaris, AIX, HP-UX, GNU/Linux, Mac OS X, and Windows platforms. It consists of a set of compiler directives, library routines, and environment variables that influence run-time behavior.

The Game of Life simulation program used the EasyX graphics library to render the graphics, using a simple black and white to represent the cell death and survival.

In this article, I designed several different parallel algorithms for Game of Life based on the OpenMP, and different methods of experimental results were analyzed, and ultimately to determine an optimal algorithm.
\section {Parallel algorithm}
The parallel algorithms are usually designed in accordance with the PCAM, Partitioning, Communication, Agglomeration and Mapping.
For the Partitioning, it is using the domain decomposition or functional decomposition approach the problem of the original calculation is divided into some small computing tasks in order to fully reveal the opportunities for parallel execution. The problem is decomposed in two forms:

Domain Decomposition, divided on the data level. Will be a big problem area is decomposed into several smaller areas, and parallel computation.

Functional Decomposition, concentrated in the level of parallel computing. Will be a big problem is decomposed into several smaller sub-problems, and parallel computation.

Different decomposition strategies affect the performance of parallel algorithms, a pilot test this influence strategy through the design of several different decomposition algorithms. Careful analysis of the game of life, to achieve parallel computation, and can only focus on the domain decomposition, because the game of life, the problem itself is small, but the deal with problem areas. Based on this analysis, the calculation part of the test program still walk through the algorithm processing, but the decomposition of the problem areas to each process in order to gain a good time performance.

Domain decomposition of the Game of Life, there are two decomposition strategy. A decomposition is the optimal decomposition by row by row or column one-dimensional decomposition. The other is a two-dimensional two-dimensional process topology structure and custom constructed type, the design of parallel algorithms is more complicated, but he is not the optimal solution.
\subsection{One-dimensional decomposition strategy}
One-dimensional decomposition strategy in two ways, by row or by column decomposition. Shown in Figure 1 and Figure 2.

Two directions decomposition of the program is feasible, Game of Life program is the array used to store the cell state, so this design, each process is responsible for calculating a bar area.
\subsection{Two-dimensional decomposition strategy}
Two-dimensional decomposition strategy is to decompose the problem areas in accordance with the size of the m * n to ensure that each process assigned to the promoter region of the same size, each process is responsible for calculating the state of the promoter region of the cell survival. Program design, the left and right margins of the region adjacent to the upper boundary and lower boundary is adjacent, in the calculation process, the neighbors of the boundary cells need to calculate the relative boundary of the cell unit. Distributed to each process in the sub-region, in order to calculate the cell survival status of the sub-regional boundaries, adjacent sub-regions corresponding to the boundary of the cell state, due to the use of an array of ways to store the cell state, cell survival status directly computing from that. Shown in Figure 3.
\begin{figure}[htbp]
\centering
\begin{minipage}[t]{0.4\textwidth}
\centering
\includegraphics{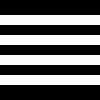}
\caption{By row decomposition}
\end{minipage}
\begin{minipage}[t]{0.4\textwidth}
\centering
\includegraphics{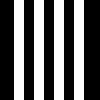}
\caption{By column decomposition}
\end{minipage}
\end{figure}

\begin{figure}
\centering
\includegraphics[width=0.4\textwidth]{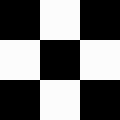}
\caption{Two-dimensional decomposition}
\label{figx}
\end{figure}
\section{Experimental Results and Performance Analysis}
For more than several different parallel algorithms based on the OpenMP,use VS2010 to write a few different test procedures, and tested in a four thread machine,due to space constraints, not to discuss the program structure.
\subsection{Experimental results}
The initial three different graphics, different algebraic results shown in Table 1.
\begin{figure}[htbp]
\centering
\begin{minipage}[t]{0.3\textwidth}
\centering
\includegraphics{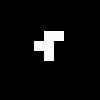}
\caption{ }
\end{minipage}
\begin{minipage}[t]{0.3\textwidth}
\centering
\includegraphics{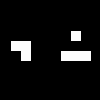}
\caption{ }
\end{minipage}
\begin{minipage}[t]{0.3\textwidth}
\centering
\includegraphics{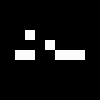}
\caption{ }
\end{minipage}
\end{figure}
\begin{table}
\begin{center}
\renewcommand{\arraystretch}{1.4}
\setlength\tabcolsep{3pt}
\begin{tabular}{llllll}
\hline\noalign{\smallskip}
$ $ & $100$ & $200$ & $500$
  & $1000$
  & $2000$\\
\noalign{\smallskip}
\hline
\noalign{\smallskip}
 Fig.4. & 121 & 120 & 202 & 156 & 124 \\
 Fig.5. & 23 & 0 & 0 & 0 & 0 \\
 Fig.6. & 76 & 169 & 267 & 393 & 256 \\
\hline
\end{tabular}
\end{center}
\caption{The total number of three graphical iteration number of the life}
\end{table}
\subsection{Performance analysis}
Parallel system speedup means for a given application: the execution speed of parallel algorithms for the execution speed of the serial algorithm to speed up many times.

\[ SP = \frac{Serial-execution-time}{Parallel-execution-time}\]

Table 2 gives figure 3 as initial graphics, the running time and speedup of different regions in different algebraic number of different processes running under the one-dimensional decomposition program.
\begin{table}
\begin{center}
\renewcommand{\arraystretch}{1.4}
\setlength\tabcolsep{3pt}
\begin{tabular}{llllll}
\hline\noalign{\smallskip}
$Solution-scale$ & $1-thread/SP$ & $2-threads/SP$ & $4-threads/SP$
  & $8 threads/SP$\\
\noalign{\smallskip}
\hline
\noalign{\smallskip}
 500 iterations(120 * 60) & 28.144/1 & 18.252/1.54 & 13.630/2.06 & 11.214/2.51 \\
 1000 iterations(120 * 60) & 54.384/1 & 34.561/1.57 & 28.398/1.91 & 26.567/2.04 \\
 2000 iterations(120 * 60) & 108.633/1 & 85.531/1.27 & 36.824/2.95 & 35.971/3.02 \\
 500 iterations(240 * 120) & 108.674/1 & 77.034/1.41 & 49.174/2.21 & 41.638/2.61 \\
 1000 iterations(240 * 120) & 218.939/1 & 125.108/1.75 & 94.846/2.41 & 67.993/3.22 \\
 2000 iterations(240 * 120) & 436.149/1 & 261.167/1.67 & 156.326/2.79 & 126.420/3.45 \\
\hline
\end{tabular}
\end{center}
\caption{Figure 3`s one-dimensional decomposition of the program testing time data (unit: s)}
\end{table}

It can be seen from Table 2, the case of the same number of threads of execution time is linearly proportional to the number of iterations; the same scale, parallel programming has a higher speed.

A conclusion can be drawn is that the time of the parallel program performance significantly better than the serial program.
\section {Conclusion}
Most computers are equipped with multi-core CPU, however, many programs still based only on single-threaded development, can not make good use of CPU resources, parallel programs can be seen through this experiment has a higher time performance relative to the serial program, but also more better use of CPU resources.

For the Game of Life problem, the two-dimensional decomposition is not the optimal solution, which can be seen in different design, parallel algorithms, different decomposition strategies on time performance of the program have a significant impact.
\end{document}